%
\let\useblackboard=\iftrue
%
%
\newfam\black
\input harvmac.tex
\def\Title#1#2{\rightline{#1}
\ifx\answ\bigans\nopagenumbers\pageno0\vskip1in%
\baselineskip 15pt plus 1pt minus 1pt
\else
\def\listrefs{\footatend\vskip 1in\immediate\closeout\rfile\writestoppt
\baselineskip=14pt\centerline{{\bf References}}\bigskip{\frenchspacing%
\parindent=20pt\escapechar=` \input
refs.tmp\vfill\eject}\nonfrenchspacing}
\pageno1\vskip.8in\fi \centerline{\titlefont #2}\vskip .5in}

\ifx\answ\bigans\def\tcbreak#1{}\else\def\tcbreak#1{\cr&{#1}}\fi
\useblackboard
\message{If you do not have msbm (blackboard bold) fonts,}
\message{change the option at the top of the tex file.}
\font\blackboard=msbm10 scaled \magstep1
\font\blackboards=msbm7
\font\blackboardss=msbm5
\textfont\black=\blackboard
\scriptfont\black=\blackboards
\scriptscriptfont\black=\blackboardss

\else

\fi
%
\def\yboxit#1#2{\vbox{\hrule height #1 \hbox{\vrule width #1
\vbox{#2}\vrule width #1 }\hrule height #1 }}
\def\fillbox#1{\hbox to #1{\vbox to #1{\vfil}\hfil}}
\def\ybox{{\lower 1.3pt \yboxit{0.4pt}{\fillbox{8pt}}\hskip-0.2pt}}
\def\np#1#2#3{Nucl. Phys. {\bf B#1} (#2) #3}
\def\pl#1#2#3{Phys. Lett. {\bf #1B} (#2) #3}

\def\physrev#1#2#3{Phys. Rev. {\bf D#1} (#2) #3}

\def\comments#1{}

\def\ZZ{{\bf Z}}
\def\RR{{\bf R}}

\def\half{{1\over 2}}

\def\CM{{\cal M}}

\def\II{\relax{I\kern-.07em I}}

\def\IZ{\relax\ifmmode\mathchoice
{\hbox{\cmss Z\kern-.4em Z}}{\hbox{\cmss Z\kern-.4em Z}}
{\lower.9pt\hbox{\cmsss Z\kern-.4em Z}}
{\lower1.2pt\hbox{\cmsss Z\kern-.4em Z}}\else{\cmss Z\kern-.4em
Z}\fi}
\def\IB{\relax{\rm I\kern-.18em B}}
\def\IC{{\relax\hbox{$\inbar\kern-.3em{\rm C}$}}}
\def\ID{\relax{\rm I\kern-.18em D}}
\def\IE{\relax{\rm I\kern-.18em E}}
\def\IF{\relax{\rm I\kern-.18em F}}
\def\IG{\relax\hbox{$\inbar\kern-.3em{\rm G}$}}
\def\IGa{\relax\hbox{${\rm I}\kern-.18em\Gamma$}}
\def\IH{\relax{\rm I\kern-.18em H}}
\def\II{\relax{\rm I\kern-.18em I}}
\def\IK{\relax{\rm I\kern-.18em K}}
\def\IP{\relax{\rm I\kern-.18em P}}

\font\cmss=cmss10 \font\cmsss=cmss10 at 7pt
\def\IR{\relax{\rm I\kern-.18em R}}

\Title{ \vbox{\baselineskip12pt\hbox{hep-th/9606017}
\hbox{RU-96-46}}}
{\vbox{
\centerline{IR Dynamics on Branes and Space-Time Geometry}}}
\centerline{Nathan Seiberg}
\smallskip
\smallskip
\centerline{Department of Physics and Astronomy}
\centerline{Rutgers University }
\centerline{Piscataway, NJ 08855-0849}
\centerline{\tt seiberg@physics.rutgers.edu}
\bigskip
\bigskip
\noindent
We consider the type I theory compactified on $T^3$.  When the D5-brane
wraps the $T^3$ it yields a D2-brane in seven dimensions.  In the
leading approximation the moduli space of vacua of the three dimensional
field theory on the brane is $T^4/\ZZ_2$.  The dual M theory description
of this theory is a compactification on K3 and our 2-brane is the
eleven dimensional 2-brane at a point in K3.  We use this fact to
conclude that strong coupling IR effects in the three dimensional theory
on the brane turn its moduli space into a K3.  This interpretation
allows us to solve various strongly coupled gauge theories in three
dimensions by identifying their Coulomb branch with a piece of a
(sometime singular) K3.

\Date{May 1996}

D-branes
\ref\clp{For a nice review see, S. Chaudhuri, C. Johnson, and J.
Polchinski, ``Notes on D-Branes,'' hep-th/9602052.}
appear to be interesting probes of space-time geometry and background
gauge fields 
\nref\dfs{U.H. Danielsson, G. Ferretti and B. Sundbor, ``D-particle
Dynamics and Bound States,'' hep-th/9603081.}%
\nref\kapo{D. Kabat and P. Pouliot, ``A Comment on Zero-brane Quantum
Mechanics,'' hep-th/9603127}%
\nref\dgdb{M.~R.~Douglas, ``Gauge Fields and D-branes,'' hep-th/9604198.}%
\nref\bds{T. Banks, M.R. Douglas and N. Seiberg, ``Probing $F$-theory
With Branes,'' hep-th/9605199}%
\nref\toap{M.~R.~Douglas, D.~Kabat, P.~Pouliot and S.~H.~Shenker,
to appear.}%
\refs{\dfs - \toap}.  In particular, motivated by observations of Sen
\ref\sen{A.~Sen, ``F-theory and Orientifolds,'' hep-th/9605150.}
the authors of \bds\ showed how the background $\tau$ parameter of  IIB
compactifications known as F theory
\ref\vafaf{C. Vafa, ``Evidence for F-theory,'' hep-th/9602022.}
can be probed.  The relevant D-brane is the D3-brane of the type II
theory, whose four dimensional world volume gauge theory has such a
$\tau$ parameter.  In the leading approximation the $\tau$ parameter is
given by the tree level, short distance value in that four dimensional
gauge theory.  The exact value of $\tau$ in F theory is given by the
long distance value in the gauge theory as determined in
\ref\swtwo{N.~Seiberg and E.~Witten, \np{B431}{1994}{484},
hep-th/9408099.}.

It is interesting that the values of the space time background fields
are given by the long distance parameters of the theory on the brane
rather than by their short distance, tree level values.  Since the brane
is only a probe of the space time fields, this observation raises an
interesting question.  How do these background 
fields adjust themselves to be compatible with the long distance
dynamics on the probes?  Are they affected by the probes or is it merely
that their values are consistent with any probes which can be used?

In \bds\ a compactification to eight dimensions was discussed.  Here
we will study lower dimensional examples. 

Consider the type I theory compactified on $T^3$ with generic $SO(32)$
Wilson lines.  Our probe will be a D5-brane which wraps the compact
$T^3$ yielding a 2-brane in seven dimensions.  The collective
coordinates of the brane are an $SU(2)$ gauge theory with some
``quarks'' 
\ref\witsmi{E. Witten, ``Small Instantons in String Theory,''
hep-th/9511030, \np{460}{1995}{541}.}.
The moduli space of vacua of the theory on the brane in the leading
approximation is parametrized by
$SU(2)$ Wilson lines.  This space is $\CM^0_{cl}=T^3/\ZZ_2$. At the
singularities of $\CM^0_{cl}$ the $SU(2)$ gauge symmetry is restored. At
generic points on $\CM^0_{cl}$ the $SU(2)$ gauge symmetry is Higgsed to
$U(1)$.  The photon in the low energy three dimensional theory can be
dualized to 
another massless boson which extends the moduli space by adding a circle
which is fibered over $T^3/\ZZ_2$.  The fiber is singular over the
singularities in the base.

The space-time $SO(32)$ Wilson lines appear as mass terms in the theory
on the brane.  Therefore, for generic values of these Wilson lines there
are no massless quarks at the singularities of $\CM^0_{cl}$.  However,
at various points on $\CM^0_{cl}$ there are massless ``electrons'' (we
distinguish them from quarks which couple to massless non-Abelian gauge
fields)  where the
$SU(2)$ Wilson line cancels the $SO(32)$ Wilson line.  At these points
the duality transformation on the photon is complicated.  However, at
the classical level $\CM^0_{cl}$ is still smooth there.

It can be helpful to perform a T duality transformation on all three
compact coordinates of the compactification.  The resulting type I'
theory is the type IIA theory compactified on $T^3$ with 8 orientifolds,
which are 6-branes and sixteen D6-branes which fill the non-compact
dimensions.  The probe is then the D2-brane of the IIA theory which is
at a point in the compact space.  The orientifold points correspond to
the enhanced $SU(2)$ gauge symmetry on the brane and the D6-branes
correspond to the points with massless electrons.

To summarize, the classical moduli space of vacua of the three
dimensional theory on the brane, $\CM_{cl}$, is a four dimensional space
whose base is $\CM^0_{cl}=T^3/\ZZ_2$ and whose fiber $S^1$ is singular
over the 8 singularities of the base.  The resulting classical moduli
space is $\CM_{cl}= T^4/\ZZ_2$.

What happens to $\CM_{cl}$ in the quantum theory?  By three dimensional
supersymmetry it has to be a hyper-Kahler manifold which is
approximately $T^4/\ZZ_2$.  If all the singularities are smoothed out,
it must become K3.  If this is the case, our 2-brane which is embedded
in seven dimensions is at a point in K3.  This is precisely the picture
implied by compactifying M theory on K3 with the eleven dimensional
2-brane identified with our 2-brane.  As a consistency check of
this proposal, note that in eleven dimensions the 2-brane is dual to the
5-brane.  Since it is at a point on the compact K3, its dual in seven
dimensions is a string obtained by wrapping the eleven dimensional
5-brane on K3.  According to the conjectured M on K3/heterotic on $T^3$
duality
\ref\wittendynamic{E. Witten, ``String Theory Dynamics in Various
Dimensions,'' hep-th/9503124, \np{443}{1995}{85}.}, 
this object is the
heterotic string
\ref\townsend{P.K. Townsend, ``String-Membrane Duality in Seven
Dimensions,'' hep-th/9504095, \pl{354}{1995}{247}.}.  
Therefore, our 2-brane is dual to the heterotic
string.  Indeed, it started its life as the 5-brane of the type I theory
which is dual to the heterotic string.

Instead of trying to prove that the moduli space becomes K3, we will
assume that this is the case.  As we said, this assumption follows from
the conjectured M on K3/heterotic on $T^3$ duality.  

We will now use this assumption as a powerful tool for studying the
dynamics of three dimensional gauge theories with $N=4$ supersymmetry.
The minimal such theory is obtained by considering a limit of the theory
on the brane without higher dimension operators.  This is the case when
the dimensionful gauge coupling $g$ in the three dimensional theory
satisfies $g^2 \ll M_{string}$.  This happens when the coupling in the
string theory is small, or equivalently, when the base $M^0_{cl}$ is
much bigger than the fiber.  Then we can study the neighborhood of
various points in $M^0_{cl}$ for various values of the moduli.

Our three dimensional theories have $N=4$ supersymmetry.  This algebra
is generated by four real supercharges $Q^i_\alpha$ ($i=1,...,4;~
\alpha=1,2$) which are spinors of the Lorentz group $SO(1,2)$.  Starting
with a renormalizable theory with massless quarks, the Lagrangian has a
global $Spin(4)_R \sim SU(2)_{R1} \times SU(2)_{R2}$ symmetry under
which the four supercharges form a vector.  This symmetry includes the
more familiar $SU(2)_R\times U(1)_R$ symmetry of the similar four
dimensional theory with $N=2$ supersymmetry.  The gauge fields are in
vector multiplets.  They include three scalars transforming as $({\bf
1,3})$ under the global symmetry and a photon.  Upon dualizing the
photon, we get another scalar transforming as $({\bf 1,1})$.  Together
with the other fields, this is a hypermultiplet.  The scalars in
ordinary hypermultiplets transform as $({\bf 2,1}) \oplus ({\bf 2,1})$
or $({\bf 3 \oplus 1,1})$.

\bigskip

\centerline{\it $U(1)$ gauge theories}

The Coulomb branch of the tree level theory is $\RR^3 \times S^1$ where
the $S^1$ factor parametrized by $\chi$ corresponds to the dual of the
photon.  The metric on this space is the obvious flat metric and the
circumference of the circle is the gauge coupling $g$.  Since the part
of the metric, which involves the circle, $g^2(d\chi)^2$, is of order
$g^2$, for consistency some higher order perturbative contributions of
the same order should be included. For example, parameterizing $\RR^3$ by
polar coordinates $\vec x=(r,\theta,\phi)$ a one loop term $c  \cos
\theta d\phi \wedge F$ ($c$ is a constant) can be generated.  After the 
duality transformation the metric becomes ${1 \over g^2 }(d \vec x)^2
+g^2 (d\chi + c\cos\theta d\phi)^2$.  Such a correction can change the
asymptotic behavior of the space.  The global $Spin(4)$ symmetry is
spontaneously broken on the moduli space away from the origin to
$SU(2)_{R1}\times U(1)$ since the scalars in the multiplet transform
like $({\bf 1,3\oplus 1})$.  The broken $SU(2)_{R2}$ factor acts on
$\RR^3$ in an obvious way and it is restored at the origin.  The circle
leads to another global $U(1)$ symmetry which is spontaneously broken.
The quantum corrections could change this picture.  In particular, in
the example above the global $SU(2)_{R2}$ acts also on the circle.
Then, $SU(2)_{R2}$ is completely broken and there is an unbroken 
global $U(1)$.

The theory without matter fields is free and the classical picture does
not change in the quantum theory.  However, with matter fields,
electrons, the theory is interacting.  The interaction is relevant at
short distance and the theory is asymptotically free.  Therefore, far
out along the flat directions, where the electrons are heavy, the moduli
space is given by the one loop corrected answer.  Comparison with the
theory on the brane can teach us about the quantum moduli space in the
strongly coupled region.

{\it $U(1)$ with one electron.}  This is the theory at generic values of
the space time moduli (i.e.\ with generic K3) when in the type I'
description the 2-brane is near one of the 16 D-branes.  In the
interior of the moduli space a singularity could develop associated with
the massless electron.  However, since our moduli space is part of a K3
at a generic point in its moduli space, which is smooth, the moduli
space of the gauge theory cannot have singularities.  We conclude that a
$U(1)$ theory with one electron does not have a singularity in its
moduli space and correspondingly, the fundamental electron is never
massless.

{\it $U(1)$ with $N_f >1$ electrons.}  In the type I' description
this theory describes the situation when $N_f$ 6-branes touch each
other and the 2-brane probe is near that point.   The theory on the
brane has an $SU(N_f)$ global symmetry while the space time theory has an
$SU(N_f)$ gauge symmetry.  Then, at the origin of the moduli space of
the theory on the brane there are $N_f$ massless electrons.  Out of that
point there is a Higgs branch.  As explained in \swtwo, on that branch
the $SU(N_f)$ global symmetry is broken to
\eqn\globsymb{SU(N_f) \rightarrow
\cases{\quad  & for $N_f=2$ \cr
U(1) & for $N_f=3$ \cr
SU(N_f-2) \times U(1) & for $N_f \ge 4$}}
The light hypermultiplets on the moduli space transform like
\eqn\lightuone{\matrix{
{\bf 1}& \quad {\rm for ~} N_f =2 \cr
{\bf 1}_1\oplus {\bf 1}_0 &\quad {\rm for ~} N_f =3 \cr
({\bf N_f-2})_1 \oplus  {\bf 1}_0 &\quad {\rm for~} N_f \ge 4.}}
As in
\ref\dbwb{M.~R.~Douglas, ``Branes within Branes,'' hep-th/9512077.}
we interpret this Higgs branch as the moduli space of an
instanton in the space time theory.  As a consistency check, note that
\globsymb\ and \lightuone\ are consistent with that interpretation.
In the quantum theory the Coulomb branch could be modified.  As
explained above, far out along the flat directions the moduli space is
given by the one loop corrected answer.  Unlike the previous cases, the
singularity on the Coulomb branch must survive.  This follows from the
existence of a Higgs branch which goes out of it.  A more precise
characterization of the singularity comes from comparing it with the
exact answer, which must be part of a K3.  Here we are at a non-generic
K3.  The enhanced $SU(N_f)$ gauge symmetry in space time is associated
with an $A_{N_f-1} $ singularity in K3.  Therefore, the moduli space
must have an $A_{N_f-1} $ singularity at the origin.  Since the short
distance theory is asymptotically free, we expect to have a different
theory at long distance.  There is no free field theory other
than the original one which leads to the Higgs branch \globsymb,
\lightuone.  Therefore, the theory at the singularity is at a
non-trivial fixed point of the renormalization group.

\bigskip

\centerline{\it $SU(2)$ gauge theories}

These theories are always interacting and asymptotically free for any
number of flavors.  The Coulomb branch is four real dimensional.  In the
classical theory it is $\RR^3\times S^1/\ZZ_2$ where the $\ZZ_2$ which
originates from the Weyl subgroup of the $SU(2)$ gauge theory acts on
both factors as reflection.  The metric on this space (away from the
singularities) is flat and the circumference of the $S^1$ factor is the
gauge coupling $g$.  Again, for consistency we should add some higher
order perturbative corrections to the metric of the same order in
$g$.  As with $U(1)$ gauge theories, the global $SU(2)_{R2}$ symmetry
acts on $\RR^3$ in an obvious way.  It is restored at the origin of the
$\RR^3$.  The perturbative quantum corrections can make $SU(2)_{R2}$ act
also on the circle.  In this case $SU(2)_{R2}$ is spontaneously broken
to a trivial group.  The global $U(1)$ symmetry which is associated with
translations of the circle is explicitly broken by instantons (the
configurations which lead to monopoles in the analogous four dimensional
theory).

{\it $SU(2)$ with no flavors.}  We find this theory at a generic
point of the space time moduli space, when in the type I' description the
2-brane probe is near one of the the orientifolds -- the origin of
$\RR^3$.  Far from there the moduli space of the gauge theory asymptotes
to the one loop corrected answer. From the map to K3, the
moduli space should be part of a smooth K3 (it is smooth because we
are at a generic point of the space time moduli space).  Therefore, the
topology of the classical space should be modified.  Such a change of
the topology of the moduli space between the classical and the quantum
theory was first found in
\ref\natimod{N. Seiberg, ``Exact Results on the Space of Vacua of Four
Dimensional SUSY Gauge Theories,'' hep-th/9402044,
\physrev{49}{1994}{6857}.}.  
In this particular example it has already been established using other
methods by Witten
\ref\wittenthreed{E. Witten, unpublished.}.
Classically the non-Abelian $SU(2)$ gauge symmetry is restored at the
origin.  However, in the quantum theory the origin is absent and the
non-Abelian gauge symmetry is never restored.  Similarly, the global
$SU(2)_{R2}$ is always broken \wittenthreed.

{\it $SU(2)$ with $N_f=1$.}  This theory is obtained on the 2-brane at a
non-generic point in the space time moduli space.  It occurs when
classically one of the 6-branes coincides with the orientifold and the
2-brane is near that point.  There is no enhanced continuous
symmetry and no Higgs branch.  As there is no enhanced continuous global
symmetry on the brane and no enhanced continuous gauge symmetry in space
time, the metric on K3 is not singular.  Therefore, the moduli space of
the three dimensional theory is smooth.  Hence, the ``W bosons'' and the
matter fields are never massless. 

{\it $SU(2)$ with $N_f>1$.}  This theory can be found when $N_f$
6-branes touch the orientifold and the 2-brane is near that point.  At 
these points there is an enhanced $SO(2N_f)$ global symmetry on the
brane and an enhanced $SO(2N_f)$ gauge symmetry in space time.  The
Higgs branch of that theory was analyzed in \swtwo.  The global symmetry
is broken as
\eqn\glosymb{SO(2N_f) \rightarrow \cases{
SU(2) & for $N_f = 2$ \cr
SO(2N_f-4)\times SU(2)  & for $N_f \ge 3$ }}
and the massless hypermultiplets transform like
\eqn\masslesq{\matrix{
{\bf 1} & {\rm for}~ N_f =2 \cr
\half ({\bf 2N_f-4},{\bf 2}) \oplus
({\bf 1},{\bf  1}) & {\rm for}~ N_f \ge 3 }}
where the $\half$ denotes half hypermultiplets in that representation.
The quantum theory is asymptotically free for any $N_f$.  Far from the
singularity the moduli space asymptotes to the one loop corrected answer
but unlike the previous case
the singularity is not smoothed out.  From the enhanced $SO(2N_f)$ gauge
symmetry in space time we know that the K3 has a $D_{N_f}$
singularity. Therefore, the moduli space of the three dimensional gauge
theory also has such a singularity.  As in \dbwb\ we interpret the
Higgs branch as the moduli space of a space time instanton. This
interpretation is consistent with \glosymb\ and \masslesq\ and with the
fact that $SO(32)$ instantons were interpreted in \witsmi\ as the Higgs
branch of the D5-brane which is our probe.  As in the $U(1)$ theory with
$N_f>1$ flavors, the theory at the singularity must be in a non-trivial
fixed point of the renormalization group.

It is trivial to extend the discussion to study other gauge groups by
using several coinciding 2-branes as a probe.  Another, less trivial,
extension of this work is to identify the three dimensional theories
whose Coulomb branch has an $E$ type singularity.  These will occur for
special values of the space time moduli.  The corresponding three
dimensional theories have global $E_{6,7,8}$ symmetries.  They are
closely related to the still mysterious theories associated with small
$E_8$ instantons
\nref\ganoha{O. Ganor and A. Hanany, ``Small E(8) Instantons and
Tensionless Noncritical Strings,'' hep-th/9602120.}%
\nref\seiwit{N. Seiberg and E. Witten, ``Comments on String Dynamics in
Six Dimensions,'' hep-th/9603003.}%
\refs{\ganoha, \seiwit}.

It is easy to repeat this analysis in lower dimensions.  Compactifying
the type I theory on $T^4$ and wrapping the D5-brane on $T^4$ we find a
string in six dimensions.  It is identified with the elementary string
of the dual type IIA theory compactified on K3.  
The moduli space of the theory on
that brane in the classical approximation is $T^4/Z_2$ \witsmi.  The
interpretation of this paper is that this space becomes a K3 (which is
smooth for generic moduli) by the quantum effects on the two dimensional
world sheet of the brane (or by some space time effect such that it is
consistent with the two dimensional dynamics).  One can
straightforwardly extend this 
discussion to compactifications to five and even four dimensions.

\medskip
\centerline{\bf Acknowledgements}
This work was supported in part by DOE grant DE-FG02-96ER40559.  We
thank T. Banks, M.  Douglas, D. Morrison, A. Sen, S. Shenker and
especially E. Witten for helpful discussions.

\bigskip

\listrefs
\end